\documentclass[twocolumn,showpacs,preprintnumbers,amsmath,amssymb
 aps,
 prb,
 lengthcheck,%
]{revtex4-1}
\usepackage{amssymb}
\usepackage{graphicx}
\usepackage{dcolumn}
\usepackage{bm}
\usepackage{hyperref}
\usepackage{makeidx}

\usepackage[english]{babel} 
\usepackage[latin1]{inputenc}
\usepackage{color}
\usepackage{graphicx}
\usepackage{amsmath}

\usepackage{latexsym}
\usepackage{amsthm}

\makeindex
\def\>{\right\rangle}
\def\<{\left\langle}
\def\be{\begin{equation}}
\def\ee{\end{equation}}
\def\ba{\begin{array}{l}}
\def\ea{\end{array}}

\begin{document}

 
\title{Electronic Hong-Ou-Mandel interferometry in two-dimensional topological insulators}

\author{D. Ferraro$^{1,2}$, C. Wahl$^{1, 2}$, J. Rech$^{1, 2}$, T. Jonckheere$^{1, 2}$, and T. Martin$^{1, 2}$ }
\affiliation{$^1$ Aix Marseille Universit\'e, CNRS, CPT, UMR 7332, 13288 Marseille, France\\
$^2$ Universit\'e de Toulon, CNRS, CPT, UMR 7332, 83957 La Garde, France}

\date{\today}

\begin{abstract}
The edge states of a two-dimensional topological insulator are characterized by their helicity, a very remarkable property which is related to the time-reversal symmetry and the topology of the underlying system. We theoretically investigate a Hong-Ou-Mandel like setup as a tool to probe it. Collisions of two electrons with the same spin show a Pauli dip, analogous to the one obtained in the integer quantum Hall case. Moreover,  the collisions between electrons of opposite spin also lead to a dip, known as $\mathbb{Z}_{2}$ dip,  which is a direct consequence of the constraints imposed by time-reversal symmetry. In contrast to the integer quantum Hall case,  the visibility of these dips is reduced by the presence of the additional edge channels, and crucially depends on the properties of the quantum point contact. As a unique feature of this system, we show the possibility of three-electron interference, which leads to a total suppression of the noise independently of the point contact configuration. This is assured by the peculiar interplay between Fermi statistics and topology. This work intends to extend the domain of applicability of electron quantum optics.

\end{abstract}

\pacs{73.23.-b, 72.70.+m, 42.50.-p}
\maketitle
\section{Introduction}

The experimental realization of on-demand electrons and holes  sources by means of driven mesoscopic capacitors \cite{Feve07, Mahe10, Buttiker93, Moskalets08} or properly designed Lorentzian voltage pulses \cite{Dubois13, Grenier13, Dubois13b} has opened the way to a new and fast developing branch of mesoscopic physics: \emph{electron quantum optics}.\cite{Grenier11b, Bocquillon13b} Its aim is to realize optics-like experiments with electrons which propagate ballistically along chiral edge channels of the integer quantum Hall (IQH) effect. The latter play the role of electron wave-guides with no backscattering, while quantum point contacts (QPCs) placed downstream act as the equivalent of beam splitters. Among the most remarkable results it is worth mentioning the electronic translation of the seminal Hanbury-Brown-Twiss \cite{Hanbury56} (HBT) and Hong-Ou-Mandel\cite{Hong87} (HOM) interferometric experiments. In the first case the fermionic nature of the electrons clearly emerges in terms of the anti-bunching between the injected electrons and the ones filling the incoming Fermi sea at finite temperature.\cite{Bocquillon12} In the second case two electronic wave-packets are injected towards the QPC on opposite edges with a tunable delay in their emission. When the emissions are perfectly synchronized one expects a suppression of the noise due to the Pauli principle which forces the electrons to emerge on opposite sides of the QPC, while for a very long delay the partition noise of two independent sources is recovered.\cite{Jonckheere12} Experimental observations validate this scenario showing the so called \emph{Pauli dip},\cite{Bocquillon13} however the suppression of the noise for zero time delay is far from being complete, which is understood as a signature of decoherence phenomena due to interaction effects between neighboring edge states.\cite{Wahl13}

 \begin{figure}[ht]
\centering
\includegraphics[scale=0.40]{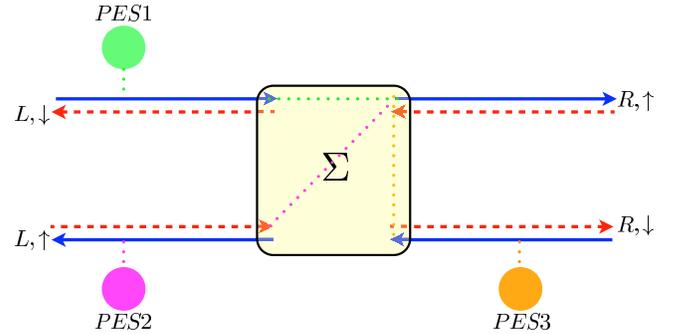}
\caption{Schematic view of a QSH bar with right-moving spin up electrons ($R\uparrow$) and left-moving spin down electrons ($L\downarrow$) along the top edge as well as left-moving spin up electrons ($L\uparrow$) and right-moving spin down electrons ($R\downarrow$) along the bottom edge. Incoming and outgoing channels are connected through a scattering region $\Sigma$ (shaded yellow square), typically given by a QPC. The possible scattering processes admitted by TRS and affecting the current in the $R\uparrow$ outgoing channel are: forward spin-preserving (dotted green), forward spin-flipping (dotted magenta) and backward spin-preserving (dotted orange). Pair-electrons sources (PES) are placed along the various edges of the sample: $PES1$ (green) inject excitations in the ($R\uparrow$) incoming and ($L\downarrow$) outgoing channels, $PES2$ (magenta) in the ($R\downarrow$) incoming and ($L\uparrow$) outgoing channels and $PES3$ (orange) in the ($L\uparrow$) incoming and ($R\downarrow$) outgoing channels.}
\label{fig2}
\end{figure}

In recent years, new states of matter showing a topological structure similar to the one of the IQH effect, but in the absence of a magnetic field, have been discovered. In particular, the two dimensional (2D) realization of these topological insulators is given by the quantum spin Hall (QSH) effect,\cite{Hasan10, Qi11} which was predicted theoretically and observed experimentally in CdTe/HgTe quantum wells \cite{Bernevig06, Konig07} and more recently also in similar structures realized with InAs/GaSb.\cite{Liu08, Knez11, Du13} This state is characterized by a gap in the bulk and peculiar gapless helical edge states 
\cite{Wu06, Qi11} in which electrons with opposite spin (or opposite total angular momentum $J_{z}$) propagate in opposite directions along the boundaries of the sample as a consequence of the strong spin-orbit coupling. While in CdTe/HgTe quantum wells these edge states are protected from backscattering by time reversal symmetry (TRS), in the new generation of 2D topological insulators the astonishing robustness of the edge states with respect to an in-plane and an out-of-plane magnetic field and the great precision of the conductance quantization seems to suggest a more general topological symmetry related to the spin Chern number.\cite{Sheng05, Yang11} 

The existence of topologically protected helical edge states in QSH naturally brings out the question about the possibility to take advantage of their peculiar spin-momentum locking properties in an electron quantum optics perspective. The first steps in this direction have been achieved very recently in Refs.~\onlinecite{Hofer13, Inhofer13}, where a characterization of the QSH equivalent of the single electron source (SES) has been discussed. Here, as long as the TRS is preserved, at each driving period a pair of electrons (holes) with opposite spin orientation is injected into the helical channels. The possibility to realize entangled electronic states by means of this kind of devices, as well as possible measurement protocols have been investigated. However, theoretical predictions concerning electron collisions of the HBT-HOM type are still lacking, and constitute our main motivation for the present work. Despite the fact that various denominations have been proposed in the literature for this kind of electronic source\cite{Hofer13, Inhofer13} from now on we will refer to it as pair-electrons source (PES) to keep in mind both the differences and the similarities with respect to the SES realized in the IQH regime. From a practical point of view, the experimental realization of a QPC in QSH systems, an essential ingredient for the proposed device, seems difficult to achieve by means of standard electrostatic gating. This is essentially due to the Klein effect, which prevents massless Dirac fermion from being confined by a potential, as also observed in graphene. \cite{Katsnelson06, Bocquillon_private} However a great experimental effort is devoted to overcome this problem by using  new generation heterostructures \cite{Knez11, Du13} or alternative techniques like the mechanical etching of the sample, raising hopes of possible relevant progress in this direction in the near future.

In this paper we consider the PES as an essential building block to realize individual electron interferometric setups (see Fig. \ref{fig2}). We focus in particular on the HBT and HOM case. We illustrate the rich phenomenology brought by the helicity of the edge channels. We observe that in the case of equal spin injection, analogous to the IQH case,\cite{Jonckheere12} a suppression of the HOM dip visibility occurs due to the presence of additional edge channels incoming and outgoing the QPC. The visibility of the dip crucially depends on the spin-flipping and spin preserving tunneling amplitude at the level of the QPC. Injections of electrons with opposite spins are also possible and lead again, remarkably enough, to HOM dips. This is not a consequence of the Pauli principle, but rather of the constraints imposed by TRS. The depth of this dip, called $\mathbb{Z}_{2}$ \emph{dip},\cite{Edge13} depends on the channels which are involved  as well as on the QPC configuration. Even more interesting is the possibility to consider three-electron injections. This configuration, which has no correspondence whatsoever in the IQH framework, is reminiscent of three-photon HOM experiments proposed in the context of quantum optics \cite{Campos00} which are within reach of nowadays multimode interference techniques.\cite{Peruzzo11} Such similarity further strengthens the deep connection between these two domains of physics. In the synchronized emission case, one observes a total suppression of the noise due to the interplay between Pauli principle and time-reversal symmetry. In the finite delay case, different behaviors are possible depending on which interference channel dominates in the transport properties. The effect of the back-flowing electrons on the functionality of the proposed setup, which is absent in the IQH case due to the chirality, is also discussed in analogy with the double SES case.\cite{Moskalets13} 

The paper is organized as follows. In Section \ref{Model} we characterize PES and QPC recalling the work of Refs. \onlinecite{Hofer13, Inhofer13}. In Section \ref{Current_Noise} we derive the expression for the outgoing current and noise in terms of the incoming ones by using a scattering matrix picture which is valid in the absence of interactions. In Section \ref{HOM} we discuss the HOM interferometer focusing on both two and three electron injections. Moreover, we provide a discussion of the possible problems associated with the helicity of the edges on the functionality of the sources. Results are summarized in section \ref{Conclusion}. An Appendix contains details about the derivation of the equations of motion of the system in presence of a QPC and the explicit form of the scattering matrix.
    
\section{Model}\label{Model}

\subsection{Mesoscopic capacitor}

\begin{figure}[ht]
\centering
\includegraphics[scale=0.50]{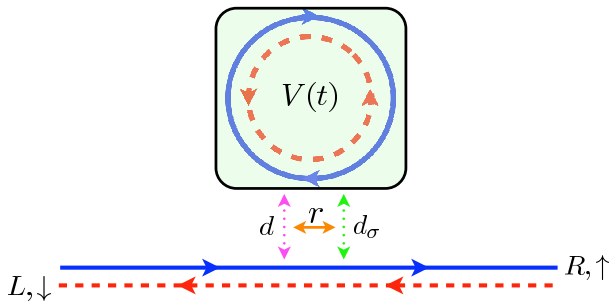}
\caption{Schematic view of a driven mesoscopic capacitor in the QSH regime. Spin up electrons (full blue lines) propagate right-moving ($R\uparrow$) along the edge bar and clockwise along the dot boundary. Spin down electrons (dashed red) propagate left-moving ($L\downarrow$) along the edge bar and anticlockwise along the dot boundary. The dot is capacitively coupled with a top gate (shaded green square) driven with a time dependent voltage $V(t)$. The dot and the edge are coupled via tunneling trough a QPC. Excitations have a probability amplitude $r$ to be reflected at the QPC (full orange arrow), a probability amplitude $d$ to be transmitted preserving their spin direction (dotted magenta arrow) and a probability amplitude $d_{\sigma}$ to be transmitted while flipping their spin (dotted green arrow).}
\label{fig1}
\end{figure}

In order to achieve electron quantum optics experiments in the QSH regime we need first to characterize a periodic source able to inject ``on-demand'' a periodic train of electrons and holes into the helical edge channels. In order to do so we can consider, in close analogy  to what was done in the 
IQH,\cite{Feve07}  a driven mesoscopic capacitor, namely a quantum dot coupled via tunneling to the edges of the QSH bar and capacitively coupled to a gate whose voltage can be periodically modulated in time (see Fig.~\ref{fig1}). This kind of setup can be realized by means of gates that separate the dot from the edge or by properly etching the sample. Note that from now on both the electrons in the dot and the ones on the QSH bar edges will be considered as freely propagating.  Moreover we will focus on the zero temperature case. 

According to the previous considerations, a dot of circumference $L$ (typically in the micrometer range) presents discrete energy levels with spacing $\Delta_{0}= h v /L$, $v$ being the Fermi velocity of the electrons (assumed equal for both the dot and the QSH bar edges for sake of simplicity).  Without loss of generality we can assume that spin up (down) electrons move clockwise (anti-clockwise) along the edges of the dot and are right-moving (left-moving) along the upper edge of the Hall bar (see Fig.~\ref{fig1}). Various possible tunneling processes, which are compatible with TRS, occur at the point contact which connects the dot with the edge of the QSH bar. In particular it is possible to have a forward transmission of electron along the edge (with amplitude probability $r$) as well as a tunneling into the dot that can both preserve (with amplitude probability $d$) or flip (with amplitude probability $d_{\sigma}$) the electron spin.\cite{Hofer13}

By keeping the two propagating channels along the QSH edge at the same chemical potential (assumed, for sake of simplicity, to be the energy reference $\mu=0$), and the dot at a relative constant voltage $V_{0}$ and subjected to a local magnetic field, the so called frozen in time scattering matrix associated with the mesoscopic capacitor is a $2\times 2$ matrix given by (up to a global phase):\cite{Hofer13}
\be
\mathcal{S}=
\left( 
\begin{matrix}
\mathcal{Y}_{-} & -d^{*} d_{\sigma} \bar{\mathcal{Z}}\\
-d d^{*}_{\sigma} \bar{\mathcal{Z}} & \mathcal{Y}_{+}\\
\end{matrix}
\right) 
\label{S_matrix}
\ee
where
\begin{align}
\mathcal{Y}_{\pm}&= -r +|d|^{2}\mathcal{Z}_{\pm}+|d_{\sigma}|^{2} \mathcal{Z}_{\mp} \\
\bar{\mathcal{Z}}&= \mathcal{Z}_{+}-\mathcal{Z}_{-}
\end{align}
with
\begin{align}
\mathcal{Z}_{\pm}(\omega)&= \int ^{+\infty}_{-\infty} dt e^{i \omega t} \sum_{q=1}^{+\infty} r^{q-1} \delta(t-q\tau_{0}) e^{-iq (\omega_{0} \tau_{0} \mp \varphi)}\nonumber\\
&= \frac{e^{i \left[(\omega-\omega_{0}) \tau_{0}\pm \varphi\right]}}{1-r e^{i \left[ (\omega-\omega_{0}) \tau_{0}\pm \varphi\right]}}.
\end{align}
in the energy (frequency) space.
 
 Note that, in the above equation, we have introduced the Josephson frequency of the dot $\omega_{0}= e V_{0}/\hbar$ ($e$ the electron charge) which represents the typical frequency associated to the excitation emitted by the dot and can be tuned by means of an external gate, the time needed by one electron to make a tour along the dot $\tau_{0}=L/v$ and the magnetic flux $\varphi$ that pierces the dot. 
 The TRS is guaranteed only in absence of magnetic field ($\varphi=0$). It is easy to note that, under this condition, the scattering matrix becomes diagonal and proportional to the identity. Because of the probability conservation 
\be
 |r|^{2}+|d|^{2}+|d_{\sigma}|^{2}=|r|^{2}+D=1,
 \label{D}
 \ee 
$D$ being the total probability for an electron to enter into the dot, the diagonal entries $\mathcal{Y}_{+}(\varphi=0)=\mathcal{Y}_{-}(\varphi=0)$ of $\mathcal{S}$ reduces to the scattering matrix already evaluated for the experimentally realized SES.\cite{Bocquillon13b, Feve07} According to the above considerations, once a periodic voltage is applied to the dot, the driven mesoscopic  capacitor realized in the QSH case behaves as two copies of the SES (one for each spin orientation). 
 
 In the SES case an optimal regime of emission is reached when the chemical potential of the edge is set precisely at the middle between two levels of the dot, a square voltage of amplitude $\Delta_{0}/e$ is applied to the dot itself and the probability of tunneling into the dot $D$ is both not too close to either zero or one.\cite{Grenier11} In this regime an electron is emitted during the first half of the period of the square wave and a hole in the second half. Analogous considerations still hold here,\cite{Inhofer13} therefore the described setup is able to inject into the two counter-propagating edge channels a pair of electrons (in the first half of the period) and a pair of holes (in the second half) with opposite orientation of the spin.  
 
 \subsection{Quantum Point Contact geometry}
 
The second step required for a proper analysis of interferometric setups in the QSH regime is the characterization of the QPC connecting the upper and the lower edges of the bar (see Fig.~\ref{fig2}). It acts as a beam splitter for electrons and holes injected along the edges by means of PES. The more general time-reversal invariant, free Hamiltonian for the system, in presence of a QPC, is given (assuming infinite edges) by \cite{Inhofer13, Dolcini11, Citro11, Romeo12}
  \be
 \mathcal{H}= \mathcal{H}_{0}+ \mathcal{H}_{sp}+\mathcal{H}_{sf}
 \ee
where
\be
\mathcal{H}_{0}= -i \hbar v\sum_{\alpha=R, L}  \sum_{\sigma= \uparrow, \downarrow} \int^{+\infty}_{-\infty} dx \xi_{\alpha} : \Psi^{\dagger}_{\alpha, \sigma}(x) \partial_{x} \Psi_{\alpha, \sigma}(x): 
\ee
is the free Hamiltonian of the two upper (right-moving spin up and left-moving spin down) and the two lower (left-moving spin up and right-moving spin down) edge channels, $\Psi_{\alpha, \sigma}(x)$ the annihilation operator for an electron of chirality $\alpha=R, L$ and spin $\sigma=\uparrow, \downarrow$, $\xi_{R/L}=\pm 1$ the chirality index and $:...:$ indicating normal ordering with respect to the Fermi sea. Focusing for simplicity on a local QPC, despite the fact that more realistic extended constrictions can also be considered,\cite{Chevallier10, Dolcetto12}  one obtains two additional contributions, namely 
\be
\mathcal{H}_{sp}= 2 \hbar v \sum_{\sigma= \uparrow, \downarrow} \gamma_{sp} \Psi^{\dagger}_{R, \sigma}(0)\Psi_{L, \sigma}(0)+h. c. 
\label{Hsp}
\ee
the spin preserving channel and
\be
\mathcal{H}_{sf}= 2 \hbar v \sum_{\alpha= R, L} \xi_{\alpha} \gamma_{sf}\Psi^{\dagger}_{\alpha, \uparrow}(0) \Psi_{\alpha, \downarrow}(0)+h.c.
\label{Hsf}
\ee
the spin flipping tunneling Hamiltonian.

According to the action of the time reversal transformation on the fermionic annihilation operators,\cite{Qi11}
\begin{align}
\mathcal{T}\Psi_{R/L, \uparrow}\mathcal{T}^{-1}&= \Psi_{L/R, \downarrow}\\
\mathcal{T}\Psi_{R/L, \downarrow}\mathcal{T}^{-1}&=- \Psi_{L/R, \uparrow},
\end{align}
the time reversal invariance of $\mathcal{H}_{sp}$ and $\mathcal{H}_{sf}$ (and consequently that of the total Hamiltonian $\mathcal{H}$) is guaranteed as long as $\gamma_{sp}$ and $\gamma_{sf}$ are real parameters. Due to the lack of direct experimental measurements of the QPC, in the following we will remain general in order to contemplate all possible experimental conditions even if, in theoretical papers, the condition $\gamma_{sp}\gtrsim \gamma_{sf}$ is typically assumed.\cite{Dolcini11, Ferraro12}

Starting from the equations of motion of the system it is possible to derive (see Appendix \ref{AppA} for a detailed discussion) the scattering matrix associated with the QPC  \cite{Inhofer13, Dolcini11, Delplace12, Edge13}
\be
\Sigma=
\left( 
\begin{matrix}
0 & \lambda_{pb}& \lambda_{ff}& \lambda_{pf}\\
\lambda_{pb}& 0 & \lambda_{pf}& \lambda_{ff}\\
\lambda^{*}_{ff} & \lambda_{pf}& 0 & \lambda_{pb}\\
\lambda_{pf} & \lambda^{*}_{ff}&  \lambda_{pb} & 0\\
\end{matrix}
\right)
\label{Sigma}
\ee
where the input and output bases are chosen as follows
\be
\left( 
\begin{matrix}
b_{L\uparrow}\\
b_{L\downarrow}\\
b_{R\uparrow}\\
b_{R\downarrow}
\end{matrix}
\right)=
\Sigma
\left(
 \begin{matrix}
a_{R\downarrow}\\
a_{R\uparrow}\\
a_{L\downarrow}\\
a_{L\uparrow}
\end{matrix}
\right).
\ee
It is worth noting that the peculiar off-diagonal form of the matrix $\Sigma$ in this basis, as well as its internal symmetry, are a direct consequence of the TRS.\cite{Dolcini11, Delplace12}  

In the above equations we introduced the parameters
\begin{align}
\lambda_{pb} &= \frac{-2 i \gamma_{sp}}{1+\gamma^{2}_{sp}+\gamma^{2}_{sf}} \\
\lambda_{ff} &= \frac{2 i \gamma_{sf}}{1+\gamma^{2}_{sp}+\gamma^{2}_{sf}} \\
\lambda_{pf} &= \frac{1-\gamma^{2}_{sp}-\gamma^{2}_{sf}}{1+\gamma^{2}_{sp}+\gamma^{2}_{sf}}
\label{eq:lambdas}
\end{align}
which represent the amplitude probabilities of spin preserving backscattering (see for example the orange dots in Fig.~\ref{fig2}), spin flipping forward scattering (magenta dots) and spin preserving forward scattering processes (green dots) respectively.\cite{Inhofer13} They satisfy the obvious conservation relation 
\be
|\lambda_{ff}|^{2}+\lambda_{pf}^{2}+|\lambda_{pb}|^{2}=1.
\label{conservation}
\ee
While this is not quite obvious from our choice of basis, one can easily verify that the scattering matrix $\Sigma$ reduces to two copies of the standard spinless form when turning off the spin flipping part of the Hamiltonian.
 
\section{Current and Noise}\label{Current_Noise}

Once both the PES and the QPC are characterized, it is possible to evaluate the current at the output of the constriction as a function of the incoming signals as well as the associated fluctuations. In the following we focus on current and auto-correlated noise in the right-moving spin up output channel $(R\uparrow)$, the other contributions can be evaluated exactly in the same way. 

The current operator we are interested in is defined as 
\be
I^{(out)}_{R\uparrow}(t)= -e v \left[:{\Psi^{\dagger}}^{(out)}_{R\uparrow}(t) \Psi^{(out)}_{R\uparrow}(t): \right]
\label{Curr_out}
\ee
and the associated noise reads 
\be
S^{(out)}_{R \uparrow, R\uparrow}(t, t')= \langle I^{(out)}_{R\uparrow}(t) I^{(out)}_{R\uparrow}(t')\rangle_{\rho}-\langle I^{(out)}_{R\uparrow}(t)\rangle_{\rho}\langle I^{(out)}_{R\uparrow}(t')\rangle_{\rho}
\ee
where the averages are taken with respect to a generic initial state described by a density matrix $\rho$.

According to the form of the scattering matrix $\Sigma$ in Eq.~(\ref{Sigma}), the outgoing electron annihilation operator can be written in terms of the incoming ones through the relation 
\be
\Psi^{(out)}_{R\uparrow}(t)= \lambda^{*}_{ff} \Psi_{R\downarrow}(t)+ \lambda_{pf} \Psi_{R\uparrow}(t)+\lambda_{pb} \Psi_{L\uparrow}(t),
\ee  
therefore the outgoing current becomes 
\begin{align}
I^{(out)}_{R\uparrow}=& -e v \left[|\lambda_{ff}|^{2} \Psi^{\dagger}_{R\downarrow}  \Psi_{R\downarrow} + \lambda_{ff}\lambda_{pf}\Psi^{\dagger}_{R\downarrow}  \Psi_{R\uparrow}\right. \nonumber \\
&+ \lambda_{ff}\lambda_{pb}\Psi^{\dagger}_{R\downarrow}  \Psi_{L\uparrow}+ \lambda_{pf}\lambda^{*}_{ff}\Psi^{\dagger}_{R\uparrow}  \Psi_{R\downarrow}\nonumber\\
&+\lambda^{2}_{pf} \Psi^{\dagger}_{R\uparrow}  \Psi_{R\uparrow}+ \lambda_{pf}\lambda_{pb}\Psi^{\dagger}_{R\uparrow}  \Psi_{L\uparrow} \nonumber\\
&+ \lambda^{*}_{pb}\lambda^{*}_{ff}\Psi^{\dagger}_{L\uparrow}  \Psi_{R\downarrow}+\lambda^{*}_{pb}\lambda_{pf}\Psi^{\dagger}_{L\uparrow}  \Psi_{R\uparrow}\nonumber\\
&+\left. |\lambda_{pb}|^{2}\Psi^{\dagger}_{L\uparrow}  \Psi_{L\uparrow}- \mathcal{G}_{F}(0)\right].
\label{current_out}
\end{align}
The last term in the above equation represents the first order coherence function,\cite{Grenier11, Grenier11b} namely the two-point Green's function, associated with the Fermi sea ($|F\rangle$)
\be
\mathcal{G}_{F}(t-t')= \langle F | \Psi^{\dagger}_{a}(t') \Psi_{a}(t) |F\rangle
\label{G_F}
\ee
which is subtracted in order to properly account for normal ordering (see Eq.~(\ref{Curr_out})). Note that the contribution which appears in Eq.~(\ref{current_out}) is given by the sum of the Fermi sea coherences of the three incoming channels ($a=R\downarrow, R\uparrow, L\uparrow$) kept at the same chemical potential, weighted by the appropriate prefactors in $\lambda$, where we explicitly took into account the additional constraint imposed by Eq.~(\ref{conservation}). It is worth mentioning that, due to the peculiar form of $\Sigma$, which is a consequence of the TRS, the presence of the $(L\downarrow)$ channel does not affect at all $I^{(out)}_{R\uparrow}$ and $S^{(out)}_{R \uparrow, R\uparrow}$.

Because of the independence of the incoming signals, the averaged output current is given by: 
\begin{align}
\langle I^{(out)}_{R\uparrow}(t)\rangle_{\rho}=& -ev \left[|\lambda_{ff}|^{2}\mathcal{G}^{(e)}_{R\downarrow}(t, t)+\lambda^{2}_{pf}\mathcal{G}^{(e)}_{R\uparrow}(t, t)\right.\nonumber \\
&+\left.|\lambda_{pb}|^{2}\mathcal{G}^{(e)}_{L\uparrow}(t, t)-\mathcal{G}_{F}(0)\right]\\
=&-ev \left[|\lambda_{ff}|^{2}\Delta\mathcal{G}^{(e)}_{R\downarrow}(t, t)+\lambda^{2}_{pf}\Delta\mathcal{G}^{(e)}_{R\uparrow}(t, t)\right.\nonumber\\
&+\left. |\lambda_{pb}|^{2}\Delta\mathcal{G}^{(e)}_{L\uparrow}(t, t)\right]
\end{align}
with 
\be
\mathcal{G}^{(e)}_{a}(t, t')= \langle \Psi^{\dagger}_{a}(t') \Psi_{a}(t) \rangle_{\rho}
\label{G_e}
\ee
the first order electronic coherence function associated with a generic incoming state \cite{Grenier11, Grenier11b} described by the density matrix $\rho$ and $\Delta \mathcal{G}^{(e)}_{a}(t, t')$ its excess with respect to the Fermi sea contribution in Eq.~(\ref{G_F}).

For further notational convenience it is useful to define, in analogy with Eq. (\ref{G_e}), also the first order hole coherence as 
\be
\mathcal{G}^{(h)}_{a}(t, t')= \langle \Psi_{a}(t') \Psi^{\dagger}_{a}(t) \rangle_{\rho}.
\label{G_h}
\ee

\begin{figure*}
\centering
\includegraphics[scale=0.55]{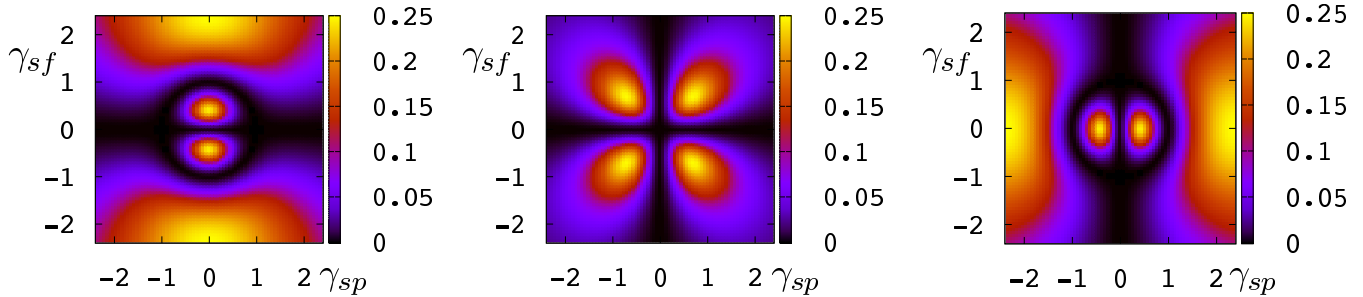}
\caption{Density plot of $\mathcal{A}$ (left), $\mathcal{B}$ (middle) and $\mathcal{C}$ (right) as a function of $\gamma_{sp}$ and $\gamma_{sf}$.}
\label{fig3}
\end{figure*}

According to Eq.~(\ref{current_out}), the outgoing noise is given by 
\begin{align}
S^{(out)}_{R\uparrow, R\uparrow}(t, t')=& |\lambda_{ff}|^{4} S_{R\downarrow, R\downarrow}(t, t') +\lambda^{4}_{pf}S_{R\uparrow, R\uparrow}(t, t')\nonumber\\
&+|\lambda_{pd}|^{4}
S_{L \uparrow, L\uparrow}(t,t')+\mathcal{Q}(t, t')
\label{S_out}
\end{align}
where the first three terms are the auto-correlated noise contributions of the incoming currents. As will be clearer in the following, transport measurement involving periodic electron sources cannot access directly the noise in the time domain, but only the zero frequency noise in Fourier space, further averaged over the emission period. It has been shown \cite{Bocquillon13b, Mahe10} that, under these conditions, the incoming noise contributions are zero and can be safely neglected. 
 
Therefore, all the interesting physics of the system is encoded in the last term of Eq.~(\ref{S_out}), which reads
\begin{align}
\mathcal{Q}(t, t')=& e^{2} v^{2} \left\{ |\lambda_{ff}|^{2} \lambda^{2}_{pf}\left[ \mathcal{G}^{(e)}_{R\downarrow}(t', t) \mathcal{G}^{(h)}_{R\uparrow}(t', t) +  (e)\leftrightarrow (h) \right]\right. \nonumber\\
&+ |\lambda_{ff}|^{2} |\lambda_{pb}|^{2}\left[ \mathcal{G}^{(e)}_{R\downarrow}(t', t) \mathcal{G}^{(h)}_{L\uparrow}(t', t) + (e)\leftrightarrow (h)  \right] \nonumber\\
&+\left.  \lambda^{2}_{pf}|\lambda_{pb}|^{2}\left[ \mathcal{G}^{(e)}_{R\uparrow}(t', t) \mathcal{G}^{(h)}_{L\uparrow}(t', t) + (e)\leftrightarrow (h)  \right]\right\},
\label{Q_G}
\end{align}
where the second term of each line is obtained by exchanging the role of electron and hole coherence functions.

It provides information about the interference between excitations from different incoming channels and generalizes what was derived in the case of the IQH.\cite{Grenier11}

In order to be as close as possible to realistic experimental situations, it is useful to introduce the new variables 
\be
t= \bar{t}+\frac{\tau}{2}, \qquad t'= \bar{t}-\frac{\tau}{2}
\ee
and to consider the quantity   
\be
Q= \int^{+\infty}_{-\infty} d\bar{t} d\tau \mathcal{Q}\left(\bar{t}+\frac{\tau}{2}, \bar{t}-\frac{\tau}{2}\right), 
\label{Q_meas}
\ee
which represents the zero frequency noise contribution (integral over $\tau$) also averaged with respect to the typical time associated with the injection of an excitation\cite{Note1} (integral over $\bar{t}$).

The measured noise contribution in Eq.~(\ref{Q_meas}) can be decomposed as 
\begin{align}
Q=& \left[ (\mathcal{A}+\mathcal{B}+\mathcal{C}) Q^{(FS)}\right.  \nonumber\\
&+ (\mathcal{A}+\mathcal{B}) Q^{(HBT)}_{R\downarrow}+ (\mathcal{A}+\mathcal{C}) Q^{(HBT)}_{R\uparrow}+ (\mathcal{B}+\mathcal{C}) Q^{(HBT)}_{L\uparrow}\nonumber\\
&+ \left. \mathcal{A} Q^{(HOM)}_{R\downarrow, R \uparrow}+\mathcal{B} Q^{(HOM)}_{R\downarrow, L \uparrow}+ \mathcal{C} Q^{(HOM)}_{R\uparrow, L \uparrow}\right]
\label{Q}
\end{align}
where we introduced the Fermi sea noise contribution \cite{Blanter00, Martin05}
\be
Q^{(FS)}= \frac{e^{2}}{\pi} \int d \bar{t} d \xi f_{\mu}(\xi) \left[1- f_{\mu}(\xi) \right]
\ee
with $f_{\mu}$ the Fermi distribution at chemical potential $\mu$ (assumed equal for all the incoming channels). The Hanbury-Brown-Twiss (HBT) contributions \cite{Bocquillon12}
\be
Q^{(HBT)}_{a}= \frac{e^{2}}{2\pi} \int d \bar{t} d \xi \Delta \mathcal{W}^{(e)}_{a}(\bar{t}, \xi) \left[1- 2f_{\mu}(\xi) \right]
\label{Q_HBT}
\ee
take into account the anti-bunching effect associated with the collision of the electrons injected in channel  $a$, against the Fermi seas of the other channels, with 
\be
\Delta \mathcal{W}^{(e)}_{a}(\bar{t}, \xi)= v \int^{+\infty}_{-\infty} d \tau e^{i \xi \tau} \Delta\mathcal{G}^{(e)}_{a}\left(\bar{t}+\frac{\tau}{2}, \bar{t}-\frac{\tau}{2} \right) 
\ee
the Wigner function associated with the excess first order electronic coherence.\cite{Ferraro13} Finally the Hong-Ou-Mandel (HOM) contributions \cite{Bocquillon13, Jonckheere12}
\be
Q^{(HOM)}_{a, b}= -\frac{e^{2}}{\pi} \int d \bar{t} d \xi \Delta \mathcal{W}^{(e)}_{a}(\bar{t}, \xi) \Delta \mathcal{W}_{b}(\bar{t}+\delta, \xi)
\label{Q_HOM}
\ee
are given by the overlap between two injected Wigner functions, where $\delta$ is the delay in the injection between the different PES in the $a$ and $b$ channels. 

The above decomposition is analogous to the one found for the IQH case, but here the phenomenology is richer and the physics depends crucially on the coefficients 
\begin{align}
\mathcal{A}=& |\lambda_{ff}|^{2} \lambda^{2}_{pf}= \frac{4 \gamma^{2}_{sf} (1-\gamma^{2}_{sp}-\gamma^{2}_{sf})^{2}}{(1+\gamma^{2}_{sp}+\gamma^{2}_{sf})^{4}}
\label{A}\\
\mathcal{B}=& |\lambda_{ff}|^{2} |\lambda_{pb}|^{2}= \frac{16 \gamma^{2}_{sf} \gamma^{2}_{sp}}{(1+\gamma^{2}_{sp}+\gamma^{2}_{sf})^{4}}\\
\mathcal{C}=&\lambda^{2}_{pf} |\lambda_{pb}|^{2}= \frac{4 \gamma^{2}_{sp} (1-\gamma^{2}_{sp}-\gamma^{2}_{sf})^{2}}{(1+\gamma^{2}_{sp}+\gamma^{2}_{sf})^{4}}
\label{C}
\end{align}
which can in principle be tuned by modifying the QPC parameters ($\gamma_{sp}$ and $\gamma_{sf}$) \cite{Romeo12, Krueckl11} as illustrated in the density plot of Fig.~\ref{fig3}. 

From these expressions, it is clear that the coefficients $\mathcal{A}$ and $\mathcal{C}$ are
simply related by the exchange of the spin-flip and spin-preserving tunneling parameters 
$\gamma_{sf}$ and $\gamma_{sp}$.

Note that in the absence of a spin-flipping term in the Hamiltonian, both coefficients $\mathcal{A}$ and $\mathcal{B}$ vanish, while $\mathcal{C}$ reduces to the product of the transmission and reflection probability of the QPC, therefore recovering the results of the IQH case. More generally, each time one of the scattering amplitudes $\lambda_{ff}$, $\lambda_{pb}$ or $\lambda_{pf}$ is zero (condition obtained for $\gamma_{sf}=0$, $\gamma_{sp}=0$ and $\gamma^{2}_{sp}+\gamma^{2}_{sf}=1$ respectively), the constraints on the system are such that only one of the parameters in Eqs. (\ref{A})-(\ref{C}) survives and the physics becomes equivalent to the one observed in the IQH (see also Eq. (\ref{Q_G})).

In the following section we discuss in detail various possible HOM interferometry experiments clarifying the expected analogies and differences with respect to what was recently observed in the IQH case.\cite{Bocquillon13}

\section{HOM interferometry}\label{HOM}

Typically in experiments it is convenient to subtract the Fermi sea contributions from $Q$ and consider only the excess noise, namely  
\be
\Delta Q= Q- (\mathcal{A}+\mathcal{B}+\mathcal{C}) Q^{(FS)},
\ee
in such a way to directly access the HBT and HOM contributions (see Eq.~(\ref{Q})) and also to eliminate possible undesired effects due to the measurement setup. 

We want now to discuss the features associated with the interference effects between electrons (holes) injected by PES that can be extracted from the measurement of $\Delta Q$. 
To simplify as much as possible the discussion, without losing any relevant physics, we can consider the emission of a single electron wave-packet from a PES in the ideal regime, the hole case can be discussed along the same lines. A generic pure incoming electron state above the Fermi sea can be written as 
\be
|\varphi_{e} \rangle= \int^{+\infty}_{-\infty} d \tau \varphi(\tau) \Psi^{\dagger}(\tau)|F\rangle
\ee 
$\varphi(\tau)$ being the electronic wave-packet in the time domain \cite{Note2} and the associated density matrix naturally reads
\be
\rho= |\varphi_{e} \rangle\langle \varphi_{e} |.
\ee

In the IQH case it has been shown \cite{Bocquillon13b, Feve07, Grenier11, Jonckheere12} that, for high enough Josephson emission frequency $\omega_{0}$, the electronic wave-packet is very well approximated by an exponential, namely  
\be
\varphi(t)\approx \sqrt {\Gamma} e^{-\frac{\Gamma}{2}t } e^{-i\omega_{0} t} \Theta(t)
\label{wp}
\ee
with \cite{Nigg98}
\be
\Gamma^{-1}= \frac{h}{\Delta_{0}} \left( \frac{1}{D}-\frac{1}{2}\right),
\ee
the parameter $D$ being defined in Eq. (\ref{D}).

Possible deviations from the above behavior, due to the presence of the Fermi sea, have been taken into account in Ref.~\onlinecite{Ferraro13}, while a full consistent treatment in terms of the Floquet scattering theory is predicted to lead to small oscillations on the time scale $\tau_{0}$ superimposed to the above envelope, that have no major effects on the physics.\cite{Jonckheere12, Moskalets13, Keeling08} The very same considerations still hold for each of the two electrons emitted by the PES along the two counter-propagating edge channels of the QSH bar. 

Assuming that all the PES are equivalent and emit wave-packets as in Eq.~(\ref{wp}), one can easily write the associated excess Wigner function that appears in both Eqs.~(\ref{Q_HBT}) and (\ref{Q_HOM}) as \cite{Ferraro13}
\begin{align}
\Delta \mathcal{W}^{(e)}(\bar{t}, \xi)=&  \int^{+\infty}_{-\infty} d \tau e^{i \xi \tau} \varphi\left( \bar{t}+\frac{\tau}{2}\right) \varphi^{*}\left( \bar{t}-\frac{\tau}{2}\right) \nonumber\\
&\approx 2 \Gamma \frac{\sin{\left[2 (\xi-\omega_{0}) \bar{t}\right]}}{\xi-\omega_{0}} e^{-\Gamma \bar{t}} \Theta(\bar{t}).
\label{wigner_wp}
\end{align}

 As long as it is possible to neglect the anti-bunching effects of the injected electron with the Fermi sea of the other channels, which seems to be a reasonable assumption in the zero temperature limit for well resolved electrons emitted at a high energy above the Fermi sea,\cite{Bocquillon12} the HBT contributions reduce to 
\be
Q^{(HBT)}_{a}\approx e^{2}
\ee
while the HOM contributions, which are no more than the overlap of identical wave-packets delayed in time,\cite{Jonckheere12} read
\be
Q^{(HOM)}_{a, b}(\delta) \approx -2 e^{2} \exp{\left(-\Gamma |\delta|\right)}.
\ee

In the following, we characterize various possible configurations of the HOM interferometer explicitly assuming the above simplifications. In particular, we consider the approximate expressions of Eqs. (\ref{wp}) and (\ref{wigner_wp}) for the wave-packet and the associated Wigner function, valid in the regime of high frequency (energy) emission for the electrons. We also neglect the effects related to the presence of a Fermi sea at zero temperature and the oscillatory behavior associated to the Floquet nature of the system.

\begin{figure*}
\centering
\includegraphics[scale=0.55]{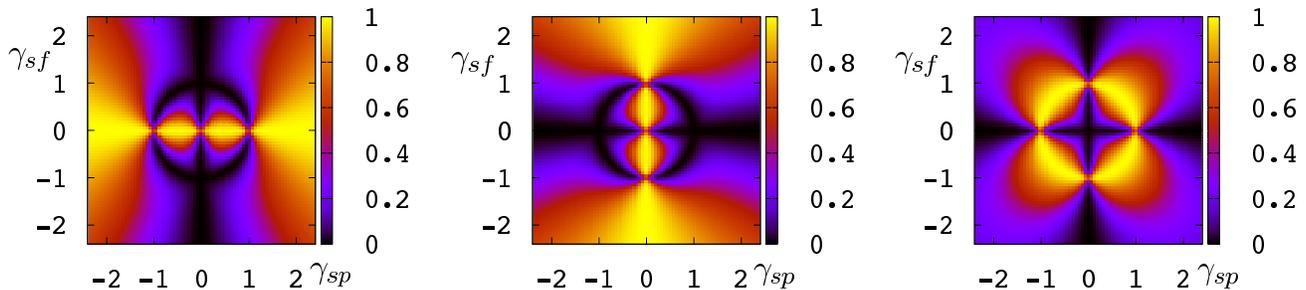}
\caption{Density plot of $\mathcal{I}$ (left), $\mathcal{J}$ (middle) and $\mathcal{K}$ (right) as a function of $\gamma_{sp}$ and $\gamma_{sf}$.}
\label{fig4}
\end{figure*}

\subsection{Two-electron collision}

As a first example of HOM interferometry experiments, we can consider the injection of electrons into the $(R\uparrow)$ and $(L\uparrow)$ incoming channels. This is realized in the setup of Fig.~\ref{fig2} when $PES1$ (green) and $PES3$ (orange) are ``on'', while $PES2$ (magenta) is ``off''. Because we are dealing with electrons with the same spin, this kind of process can be considered as the direct transposition in the QSH framework of the IQH case without interaction. According to this, a good quantity to study is given by the adimensional ratio between the measured excess noise when two sources emit together with finite delay $\delta$ and the sum of the HBT contributions from the same sources taken separately, namely 
\begin{widetext}
\be
q^{(2)}_{R\uparrow, L\uparrow}(\delta)= \frac{(\mathcal{A}+\mathcal{C}) Q^{(HBT)}_{R \uparrow}+(\mathcal{B}+\mathcal{C}) Q^{(HBT)}_{L \uparrow}+\mathcal{C} Q^{(HOM)}_{R\uparrow, L\uparrow}(\delta)}{(\mathcal{A}+\mathcal{C}) Q^{(HBT)}_{R \uparrow}+(\mathcal{B}+\mathcal{C}) Q^{(HBT)}_{L \uparrow}} \approx 1- \mathcal{I}e^{-\Gamma |\delta|}
\label{q2_upup}
\ee  
\end{widetext}
where 
\be
\mathcal{I}= \frac{2\mathcal{C}}{\mathcal{A}+\mathcal{B}+2\mathcal{C}}
\ee
is a visibility factor, whose behavior as a function of the QPC parameters is shown in Fig.~\ref{fig4} (left panel).

\begin{figure}[ht]
\centering
\includegraphics[scale=0.25]{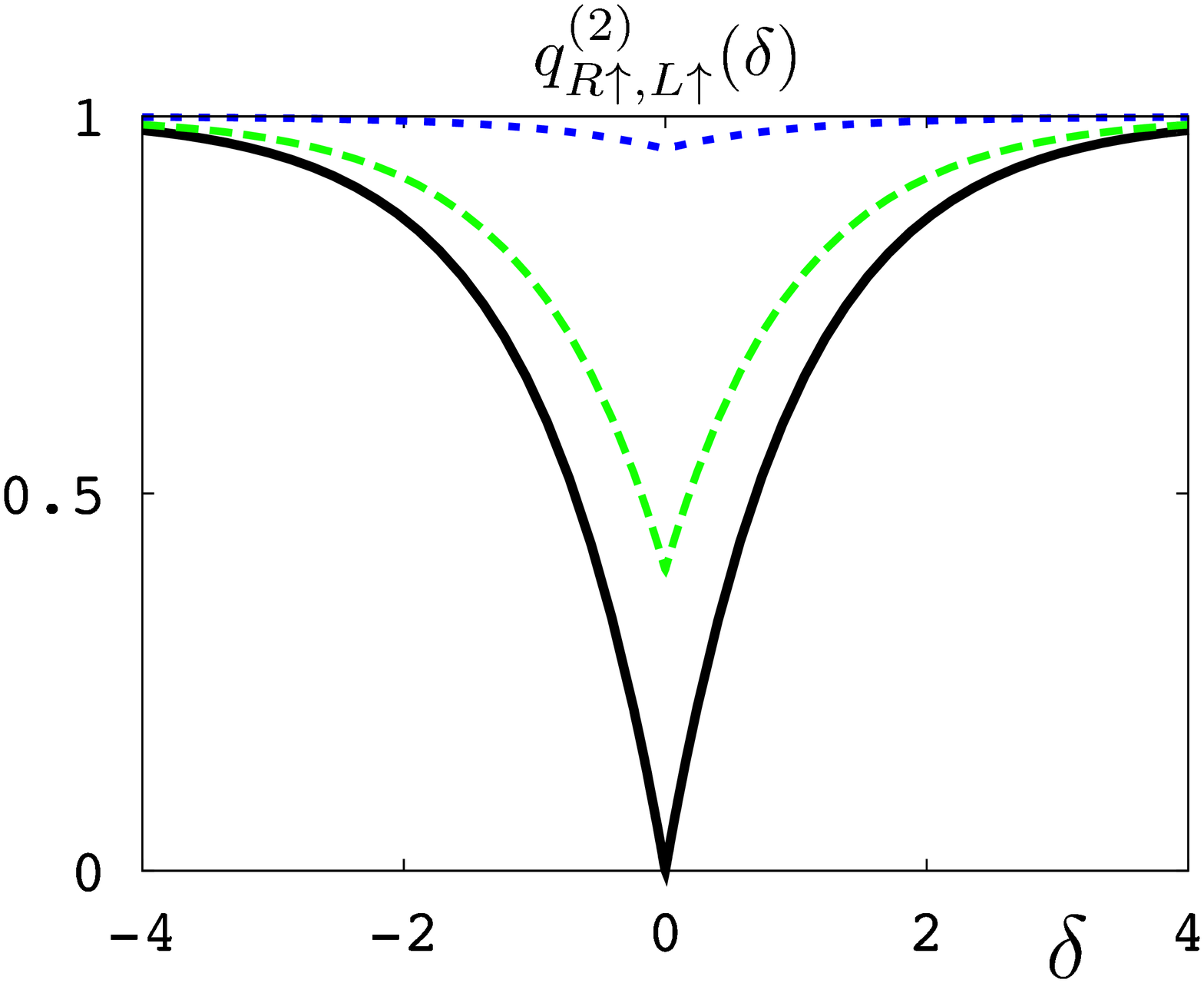}
\caption{Behavior of $q^{(2)}_{R\uparrow, L\uparrow}(\delta)$ as a function of the delay $\delta$ (in units of $\Gamma^{-1}$) for different values of the QPC spin-flipping and spin preserving amplitudes: $\gamma_{sp}=2, \gamma_{sf}=0$ (full black), $\gamma_{sp}=2, \gamma_{sf}=1.5$ (dashed green) and $\gamma_{sp}=1, \gamma_{sf}=0.3$ (dotted blue).}
\label{fig5}
\end{figure}

Similarly to the IQH case,\cite{Jonckheere12} Eq.~(\ref{q2_upup}) indicates the appearance of a dip in the noise when the electrons reach the QPC with a delay comparable with the typical extension in time of the wave-packet ($\Gamma|\delta| \lesssim 1$) and whose exponential form is reminiscent of Eq.~(\ref{wp}). This \emph{Pauli dip} is a consequence of the fermionic statistics of the electron.\cite{Bocquillon13} 
However, in the case considered here, the amplitude of the dip is reduced compared to what is observed for the integer quantum Hall effect without interaction.\cite{Jonckheere12} This reduced visibility can be interpreted as a direct consequence of the additional channels which are coupled at the QPC. \cite{Rizzo13}
Indeed, more channels for the electrons to scatter into means an increased noise associated with partitioning at the QPC (corresponding to the terms in $\mathcal{A}$ and $\mathcal{B}$ in Eq.~\eqref{q2_upup}).
This increase generally cannot be compensated for by the noise reduction related to the Pauli principle through the HOM contribution, ultimately leading to a reduced dip.
This effect can be negligible, with a visibility $\mathcal{I}\approx 1$ (see Fig.~\ref{fig5} full black and dashed green curve), or conversely very relevant, with $\mathcal{I}\approx 0$ (see Fig.~\ref{fig5} dotted blue curve), and crucially depends on the intensity of $\gamma_{sp}$ and $\gamma_{sf}$ (see Fig.~\ref{fig4}). In the absence of spin flipping ($\gamma_{sf}=0$ and consequently $\mathcal{A}=\mathcal{B}=0$), we recover the result of the IQH ($\mathcal{I}=1$) as a consequence of the complete decoupling of the QSH system into two IQH-like states with opposite spin orientations. 
It is worth remarking that this suppression of the dip visibility differs from what was recently observed\cite{Bocquillon13, Wahl13} in experiments carried out in the IQH effect at filling factor $\nu=2$, where the loss of contrast is a consequence of inter-channel Coulomb repulsion. While both setups involve multiple channels, here it occurs in a non-interacting regime, and arises from multiple scattering processes at the QPC.

Due to the helical properties of the edge states, new two-electron interference processes are possible involving particles with opposite spin. This can first be achieved with electrons of the same chirality, namely 
\be
q^{(2)}_{R\downarrow, R\uparrow}(\delta)\approx 1- \mathcal{J} e^{-\Gamma |\delta|}
\label{q2_RR}
\ee
where $PES1$ (green) and $PES2$ (magenta) are ``on'' while $PES3$ (orange) is ``off'' (see Fig.~\ref{fig2}) and we introduced another visibility
\be
\mathcal{J}= \frac{2\mathcal{A}}{2\mathcal{A}+\mathcal{B}+\mathcal{C}}. 
\ee
It is transparent from the Hamiltonian describing the QPC, Eqs.~\eqref{Hsp} and \eqref{Hsf}, that the collision process involving $(R\uparrow)$ and $(L\uparrow)$ electrons, and the one involving $(R\uparrow)$ and $(R\downarrow)$ electrons, are related to one another under the exchange of the spin-preserving and spin-flipping contributions. 
This particular symmetry ensures a similar link between the coefficients $\mathcal{A}$ and $\mathcal{C}$, which ricochets onto the visibility factors $\mathcal{I}$ and $\mathcal{J}$, thus also connected under the exchange of $\gamma_{sp}$ and $\gamma_{sf}$ (see the middle panel of Fig.~\ref{fig4}).

As a consequence, the maximum visibility for the HOM interferences, corresponding to $\mathcal{J}=1$,
is obtained by taking $\gamma_{sp}=0$, which leads to $\mathcal{B}=\mathcal{C}=0$. Here, the absence of spin-preserving tunneling implies that the amplitude $\lambda_{pb}$ of the spin preserving backscattering is zero (see Eq.~(\ref{eq:lambdas})).

\begin{figure}[ht]
\centering
\includegraphics[scale=0.25]{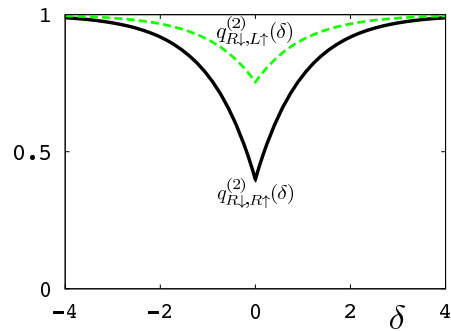}
\caption{Behavior of $q^{(2)}_{R\downarrow, R\uparrow}(\delta)$ (full black) and $q^{(2)}_{R\downarrow, L\uparrow}(\delta)$ (dashed green) as a function of the delay $\delta$ (in units of $\Gamma^{-1}$). Parameters are $\gamma_{sp}=\gamma_{sf}=2$.}
\label{fig6}
\end{figure}

Even more interesting is the possibility to probe the interference of electrons with both opposite spin and chirality
\be
q^{(2)}_{R\downarrow, L\uparrow}(\delta)\approx 1-\mathcal{K} e^{-\Gamma |\delta|}
\label{q2_downup}
\ee
with $PES2$ (magenta) and $PES3$ (orange) turned ``on'' while $PES1$ (green) is ``off'' (see Fig.~\ref{fig2}) and 
\be
\mathcal{K}= \frac{2\mathcal{B}}{\mathcal{A}+2\mathcal{B}+\mathcal{C}}.
\ee
Here again it is possible to have a maximum visibility for the HOM interferences, 
with $\mathcal{K}=1$. This requires $\mathcal{A}=\mathcal{C}=0$, which is obtained for 
$\gamma_{sp}^2 + \gamma_{sf}^2 =1$. The region of maximum visibility is thus a circle
of radius 1 in the ($\gamma_{sp},\gamma_{sf}$) plane, as can be seen on the right panel
of Fig.~\ref{fig4}.
 When this condition is satisfied the spin preserving forward
scattering amplitude $\lambda_{pf}$ is zero (see Eq.~(\ref{eq:lambdas})).
 
 In the three different two-electron collision configurations which
 we have considered (Eqs.~(\ref{q2_upup}), (\ref{q2_RR}) and (\ref{q2_downup})),
  the maximum visibility is always obtained by choosing 
 $\gamma_{sp}$ and $\gamma_{sf}$ such that one of the scattering amplitudes 
 (respectively $\lambda_{ff}$, $\lambda_{pb}$ or $\lambda_{pf}$) is zero. When this 
 condition is satisfied (namely for $\gamma_{sf}=0$, $\gamma_{sp}=0$ and $\gamma^{2}_{sp}+\gamma^{2}_{sf}=1$ respectively) only two possible exit channels remain available for the two electrons
 and we recover a noiseless output for synchronized electrons, in full analogy with the IQH case. 

The possibility to have noise suppression also for interferometers involving electrons with opposite spin is related to the peculiar constraints imposed by TRS and charge conservation in the QSH system. This has been remarked for the first time in Ref.~\onlinecite{Edge13} for the continuous current case. This phenomenon, known as $\mathbb{Z}_{2}$ \emph{dip}, has also been predicted by Inhofer \emph{et al.} in Ref.~\onlinecite{Inhofer13} in the case of periodic injections through PES, however for a specific range of parameters ($\gamma_{sp}=0$, $\gamma_{sf}\neq0$), which seems difficult to be obtained in real experiments.\cite{Dolcini11} The analysis reported here proves more general, as it not only recovers this previous result, but also predicts the existence of a reduced dip in the more general, experimentally relevant situation, accounting for the decrease in visibility  (encoded in the coefficients $\mathcal{J}$ and $\mathcal{K}$) due to the multiple scattering processes at the QPC, as shown in Fig.~\ref{fig6}. 

\subsection{Three-electron collision}

Let us consider finally another configuration which is unique to the quantum spin Hall effect and which can be seen as the electron quantum optics translation of three-photon HOM experiments.\cite{Campos00} Here, all the PES of Fig.~\ref{fig2} are switched on, with possible relative delays in the emissions.\cite{Note3} To fix the notation we  label $\delta_{1}$ the time interval between the injections $(R\downarrow-R\uparrow)$, and  $\delta_{2}$ the one for $(R\downarrow-L\uparrow)$. Consequently $(\delta_{2}-\delta_{1})$ corresponds to the delay for $(R\uparrow-L\uparrow)$. As a natural extension of previous calculations, it is possible to define an adimensional noise
\begin{widetext}
\begin{align}
q^{(3)}(\delta_{1}, \delta_{2}) &= \frac{(\mathcal{A}+\mathcal{B}) Q^{(HBT)}_{R \downarrow}+(\mathcal{A}+\mathcal{C}) Q^{(HBT)}_{R \uparrow}+(\mathcal{B}+\mathcal{C}) Q^{(HBT)}_{L \uparrow}+\mathcal{A} Q^{(HOM)}_{R\downarrow, R\uparrow}(\delta_{1})+\mathcal{B} Q^{(HOM)}_{R\downarrow, L\uparrow}(\delta_{2})+\mathcal{C} Q^{(HOM)}_{R\uparrow, L\uparrow}(\delta_{2}-\delta_{1})}{(\mathcal{A}+\mathcal{B}) Q^{(HBT)}_{R \downarrow}+(\mathcal{A}+\mathcal{C}) Q^{(HBT)}_{R \uparrow}+(\mathcal{B}+\mathcal{C}) Q^{(HBT)}_{L \uparrow}}\nonumber\\
&\approx  1- \frac{\mathcal{A}}{\mathcal{A}+\mathcal{B}+\mathcal{C}} e^{-\Gamma |\delta_{1}|}-\frac{\mathcal{B}}{\mathcal{A}+\mathcal{B}+\mathcal{C}} e^{-\Gamma |\delta_{2}|}-\frac{\mathcal{C}}{\mathcal{A}+\mathcal{B}+\mathcal{C}} e^{-\Gamma |\delta_{2}-\delta_{1}|}.
\end{align}  
\end{widetext}

Note that, for synchronized injections on the three channels, one has 
\be
q^{(3)}(\delta_{1}=0, \delta_{2}=0)=0
\ee
\emph{independently of the characteristics of the QPC}. This total suppression of the noise is a very remarkable feature of helical systems and generalizes both the \emph{Pauli dip} and the $\mathbb{Z}_{2}$ \emph{dip} structure. It depends on the extremely peculiar interplay between the fermionic statistics and the TRS in the helical edge states.  

This is better illustrated with the help of a simple calculation. Consider, as a simplified version of the three-electron injection, the following input state $a_{R\uparrow}^\dagger a_{R\downarrow}^\dagger a_{L\uparrow}^\dagger | F \rangle$, corresponding to the creation of three electrons simultaneously in the three input channels $(R\uparrow)$, $(R\downarrow)$ and $(L\uparrow)$. Using the expression for the scattering matrix $\Sigma$, Eq.~\eqref{Sigma}, one can rewrite this in terms of the outgoing electronic creation operators as:
\begin{align}
a_{R\uparrow}^\dagger a_{R\downarrow}^\dagger a_{L\uparrow}^\dagger  | F \rangle =& \left( \lambda_{pb}^* b_{L\uparrow}^\dagger + \lambda_{pf} b_{R\uparrow}^\dagger + \lambda_{ff}^* b_{R\downarrow}^\dagger \right) \nonumber \\
& \times \left( \lambda_{pb}^* b_{L\downarrow}^\dagger + \lambda_{ff}^* b_{R\uparrow}^\dagger + \lambda_{pf} b_{R\downarrow}^\dagger \right) \nonumber \\
& \times \left( \lambda_{pf} b_{L\uparrow}^\dagger + \lambda_{ff} b_{L\downarrow}^\dagger + \lambda_{pb}^* b_{R\uparrow}^\dagger \right) | F \rangle
\end{align}
Note that the TRS ensures that each input operator $a_{\alpha\sigma}^\dagger$ is expressed in terms of only three out of the four possible output operators $b_{\alpha\sigma}^\dagger$. 

Expanding this expression, using the unitarity of the scattering matrix and invoking the Pauli principle to get rid of all squared operators, we are left with
\begin{align}
a_{R\uparrow}^\dagger a_{R\downarrow}^\dagger a_{L\uparrow}^\dagger | F \rangle =& \left( \lambda_{pb} b_{L\uparrow}^\dagger  b_{L\downarrow}^\dagger   
+ \lambda_{pf} b_{R\downarrow}^\dagger  b_{L\uparrow}^\dagger   
\right. \nonumber \\
&  \left. + \lambda_{ff} b_{R\downarrow}^\dagger  b_{L\downarrow}^\dagger  \right) 
b_{R\uparrow}^\dagger| F \rangle
\end{align}
This can thus be viewed as the superposition of three possible outgoing states, which all involve the creation of an electron in the $(R\uparrow)$ output channel, precisely the one where we measure current and noise. This means, in particular, that there are no current fluctuations in this channel as it is always populated, no matter the final outcome of the scattering process at the QPC. In other words, there cannot be any partition noise in the channel of interest for this three-electron injection, a feature which we could attribute to the effect of both  TRS and the Pauli principle.

\begin{figure}[h]
\centering
\includegraphics[scale=0.55]{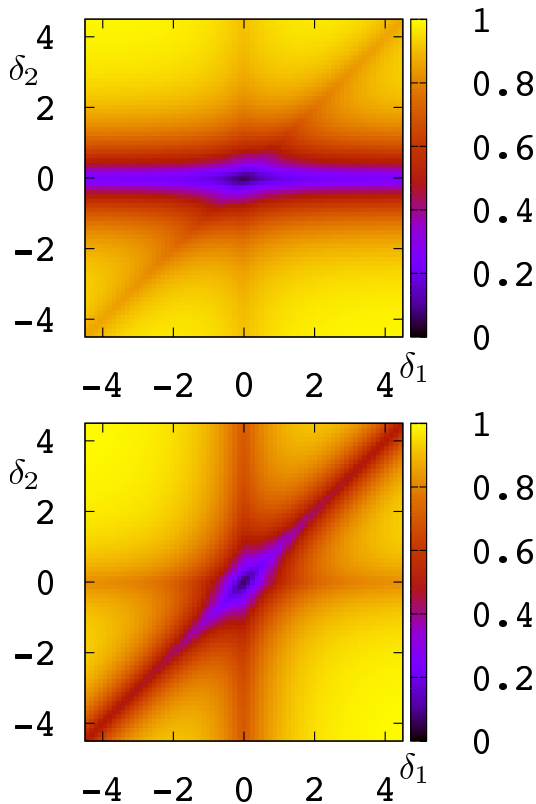}
\caption{Density plot of $q^{(3)}(\delta_{1}, \delta_{2})$ as a function of $\delta_{1}$ and $\delta_{2}$ (in units of $\Gamma^{-1}$) for $\gamma_{sp}=1, \gamma_{sf}=0.8$ (top) and $\gamma_{sp}=2, \gamma_{sf}=1.5$ (bottom).}
\label{fig7}
\end{figure}

For non synchronized injections one obtains an extremely rich phenomenology depending on the QPC microscopic parameters $\gamma_{sp}$ and $\gamma_{sf}$. For example, by changing the properties of the QPC, one can move from a situation where the noise suppression is roughly independent of $\delta_{1}$ as shown in Fig.~\ref{fig7} (top panel), indicating a dominance of the ($R\downarrow, L\uparrow$) interference process, to one in which the HOM dip is more pronounced for $\delta_{1}\approx \delta_{2}$, a signature of an important contribution of equal spin injection ($R\uparrow, L\uparrow$) shown in Fig.~\ref{fig7} (bottom panel). These considerations open the way to investigate in a unique setup different interference configurations, by tuning the QPC parameters. In a complementary way this kind of HOM measurements represent a useful tool in order to characterize the QPC, providing interesting information about the relative importance between the spin-preserving and the spin-flipping microscopic tunneling amplitudes.

\subsection{Effects of back-flowing electrons} 

In order to better characterize our setup in view of an experimental realization of the previously described HOM like interferometers using QSH edge states, it is worth commenting about possible drawbacks related to helical propagation of the excitations along the edges. Differently from what happens in the chiral case, here electrons scattered at the QPC can back-flow into another PES altering in principle its functionality. As an example of this we can consider an electron injected in the $(L\uparrow)$ incoming channel by $PES3$ (orange) shown in Fig.~\ref{fig2}. In general it has a non zero probability to be scattered in the $(L\uparrow)$ outgoing channel. Once arrived in correspondence to $PES2$ (magenta) it can affect the emission of an excitation in the same channel, and consequently the one in the $(R\downarrow)$ incoming channel, in various possible ways: \emph{i)} the arrival of the electron (hole) is not synchronized with the emission of the excitations from $PES2$, \emph{ ii)} the arrival of the electron (hole) is synchronized with the emission of a hole (electron), \emph{iii)} the arrival of an electron (hole) is synchronized with the emission of an electron (hole). In order to discuss these situations we can recover what was done in Ref.~\onlinecite{Moskalets13} for the physically equivalent situation of two SES placed along the same chiral channel of the IQH. As expected the condition \emph{i)} doesn't affect the functionality of $PES2$ due to the independence between arrival and emission. Situation \emph{ii)} is more troublesome because the electron (hole) could in principle be reabsorbed by $PES2$ leading to the annihilation of the emitted hole (electron). However it is possible to show that, for asymmetric wave-packets in the time domain (not time reversal invariant) as the ones considered here (see Eq.~(\ref{wp})), this resorption is forbidden and the functionality of $PES2$ is preserved. This could not be true away from the optimal regime and in particular in the adiabatic regime, where a low frequency sinusoidal drive is applied to the dot \cite{Hofer13} and the emitted wave-packet is Lorentzian (thus symmetric) in time. Only the final situation \emph{iii)} is truly problematic. When the arrival of an electron (hole) is synchronized with the pair injection, the injected electron is emitted at a higher energy because of the Pauli principle. This can affect the previous discussion (in particular the parameters of the emitted wave-packet are modified). Therefore the setup needs to be engineered in such a way to minimize these effects, namely reduce as much as possible the overlap between the wave-packet of the arriving and of the injected electron. Note that the discussion summarized here can be rigorously derived in a full consistent treatment based on the Floquet scattering theory.\cite{Moskalets13} 

\section{Conclusion}\label{Conclusion}

In this paper we considered the possibility to realize electron quantum optics experiments in the framework of new materials relevant to modern
condensed matter physics called topological insulators, which have attracted a lot of attention recently, both theoretically and experimentally. 
We focused on their two-dimensional version exhibiting the QSH effect. By means of PES which inject pairs or electrons and holes into the helical edge channels it is possible to realize HBT and HOM interferometers. Focusing on the latter case we have observed a very rich and interesting phenomenology related to the peculiar spin-momentum locking of the electrons propagating along the edges. In the case of a two-electron injection (with two sources ``on'' and one source ``off'') it is possible to realize either an interference between electrons with the same spin (reminiscent of the HOM dip observed in the non interacting IQH case but with a visibility reduced by the presence of additional channels) or an interference between electrons with opposite spin, where the observed dip is due to the constraints imposed by the topological structure of the edges protected by TRS. The presented setup also allows to realize three-electron injection which is characterized by a total suppression of the noise in the case of perfect synchronization and shows different possible behaviors depending on the QPC spin-preserving and spin-flipping tunneling amplitudes, which can modify the relevance of the different interference contributions. Such
three-electron interference phenomenon bears a three-photon equivalent in the context of quantum optics, and have so far eluded investigation in an electronic condensed matter setting. We have pointed out that our prediction on HOM interferometry could in principle be used to characterize the QPC in actual experiments, providing interesting information about the relative importance between the spin-preserving and the spin-flipping microscopic tunneling amplitudes.

Possible extensions of this work include the effect of finite temperature, which should not modify drastically our present results as long as the injected electrons are well resolved above the Fermi sea. However, finite temperature effects should be of importance when studying 
collisions between injected electrons and injected holes, as was uncovered in the IQH case.\cite{Jonckheere12} 
Another issue concerns the effect of interactions, which have been neglected here, but which are known to operate in two 
dimensional systems with edge channels. Interaction effects could in principle be taken into account in terms of the so called 
helical Luttinger liquid picture for both the edges \cite{Strom09, Hou09} and the dot. \cite{Dolcetto13} On the basis 
of a recent work of some of the authors\cite{Wahl13} on IQH bars where edge channels co-propagate however, 
we suspect that Coulomb interaction among the counter-propagating edges 
could lead to a further reduction of the visibilities of the \emph{Pauli} and the $\mathbb{Z}_{2}$ \emph{dips}  in our two electron collision predictions, 
and give rise to a non vanishing of the three particle dip.  

\section*{Acknowledgements}
The authors acknowledge the support of ANR-2010-BLANC-0412 (``1 shot''). This work has been carried out in the framework of the Labex Archim\`ede (ANR-11-LABX-0033) and of the A*MIDEX project (ANR-11-IDEX-0001-02), funded by the ``Investissements d'Avenir'' French Government program managed by the French National Research Agency (ANR).

\appendix 

\section{Equations of motion}\label{AppA}

In this Appendix we summarize the well known results about the derivation of the equations of motion of the system described by the Hamiltonian $\mathcal{H}= \mathcal{H}_{0}+\mathcal{H}_{sp}+\mathcal{H}_{sf}$ and the scattering matrix associated to the QPC. 

Using the Heisenberg evolution equation 
\be
i \hbar \partial_{t}\Psi_{\alpha, \sigma} = \left[ \Psi_{\alpha, \sigma}, \mathcal{H}\right]
\ee
one has (the dependence on time $t$ and space $x$ of the operators is implied for notational convenience)
\begin{align}
i \partial_{t} \Psi_{R\uparrow}&=& -i v \partial_{x} \Psi_{R\uparrow}+ 2 v\delta(x)\left(\gamma_{sp} \Psi_{L\uparrow}+ \gamma_{sf} \Psi_{R\downarrow} \right)\nonumber \\
i \partial_{t} \Psi_{R\downarrow}&=& -i v \partial_{x} \Psi_{R\downarrow}+ 2 v\delta(x)\left(\gamma_{sp} \Psi_{L\downarrow}+ \gamma_{sf} \Psi_{R\uparrow} \right)\nonumber \\
i \partial_{t} \Psi_{L\uparrow}&=& +i v \partial_{x} \Psi_{L\uparrow}+ 2 v\delta(x)\left(\gamma_{sp} \Psi_{R\uparrow}- \gamma_{sf} \Psi_{L\downarrow} \right)\nonumber \\
i \partial_{t} \Psi_{L\downarrow}&=& +i v \partial_{x} \Psi_{L\downarrow}+ 2 v\delta(x)\left(\gamma_{sp} \Psi_{R\downarrow}- \gamma_{sf} \Psi_{L\uparrow} \right).\nonumber\\
\end{align}

This can be solved in terms of the plane-waves ansatz for the first quantized electronic wave-functions
\be
\psi_{R, \sigma}=\frac{e^{-i\frac{E}{\hbar} t}}{\sqrt{h v }} 
\left\{ 
\begin{matrix}
 a_{R, \sigma} e^{i k_{E} x}& x<0\\
 b_{R, \sigma} e^{i k_{E} x}& x>0\\
\end{matrix}
\right. 
\label{S_matrix}
\ee
and
\be
\psi_{L, \sigma}=\frac{e^{-i\frac{E}{\hbar} t}}{\sqrt{h v }} 
\left\{ 
\begin{matrix}
 b_{L, \sigma} e^{-i k_{E} x}& x<0\\
 a_{L, \sigma} e^{-i k_{E} x}& x>0\\
\end{matrix}
\right.. 
\label{S_matrix}
\ee
It is now possible to integrate the equations of motion in the infinitesimal interval $\left[-\epsilon, +\epsilon\right]$ ($\epsilon\rightarrow 0^{+}$). Due to the fact that we are dealing with Dirac equation (first order in the space derivative) with delta-like potentials, one needs to use the regularization
\be
f(0)=\frac{1}{2}\left[f(0^{-})+f(0^{+}) \right]
\ee
for the terms associated to the delta-like contributions.\cite{Sutherland81, Roy93} According to this, one obtains the set of equations
\begin{align}
-i \left(b_{R\uparrow}-a_{R\uparrow} \right)+\gamma_{sp} \left( a_{L\uparrow}+b_{L\uparrow}\right)+\gamma_{sf}\left(b_{R\downarrow}+a_{R\downarrow} \right)&=&0\nonumber\\ 
-i \left(b_{R\downarrow}-a_{R\downarrow} \right)+\gamma_{sp} \left( a_{L\downarrow}+b_{L\downarrow}\right)+\gamma_{sf}\left(b_{R\uparrow}+a_{R\uparrow} \right)&=&0\nonumber\\
-i \left(b_{L\uparrow}-a_{L\uparrow} \right)+\gamma_{sp} \left( a_{R\uparrow}+b_{R\uparrow}\right)-\gamma_{sf}\left(b_{L\downarrow}+a_{L\downarrow} \right)&=&0\nonumber\\  
-i \left(b_{L\downarrow}-a_{L\downarrow} \right)+\gamma_{sp} \left( a_{R\downarrow}+b_{R\downarrow}\right)-\gamma_{sf}\left(b_{L\uparrow}+a_{L\uparrow} \right)&=&0\nonumber\\
\end{align}
that can be solved in terms of the outgoing states in the form
\be
\left( 
\begin{matrix}
b_{L\uparrow}\\
b_{L\downarrow}\\
b_{R\uparrow}\\
b_{R\downarrow}
\end{matrix}
\right)=
\Sigma
\left(
 \begin{matrix}
a_{R\downarrow}\\
a_{R\uparrow}\\
a_{L\downarrow}\\
a_{L\uparrow}
\end{matrix}
\right)
\label{Sigma_matrix}
\ee
where
\be
\Sigma=
\left( 
\begin{matrix}
0 & \lambda_{pb}& \lambda_{ff}& \lambda_{pf}\\
\lambda_{pb}& 0 & \lambda_{pf}& \lambda_{ff}\\
\lambda^{*}_{ff} & \lambda_{pf}& 0 & \lambda_{pb}\\
\lambda_{pf} & \lambda^{*}_{ff}&  \lambda_{pb} & 0\\
\end{matrix}
\right) 
\ee
with 
\begin{align}
\lambda_{pb}&= \frac{-2 i \gamma_{sp}}{1+\gamma^{2}_{sp}+\gamma^{2}_{sf}} \\ 
\lambda_{ff}&= \frac{2 i \gamma_{sf}}{1+\gamma^{2}_{sp}+\gamma^{2}_{sf}} \\
\lambda_{pf}&= \frac{1-\gamma^{2}_{sp}-\gamma^{2}_{sf}}{1+\gamma^{2}_{sp}+\gamma^{2}_{sf}}.
\end{align}

\end{document}